\documentclass[pdftex,epjc3]{svjour3}          

\RequirePackage[T1]{fontenc}

\smartqed  

\RequirePackage{graphicx}
\RequirePackage{mathptmx}      
\RequirePackage{flushend}
\RequirePackage[numbers,sort&compress]{natbib}
\RequirePackage[colorlinks,citecolor=blue,urlcolor=blue,linkcolor=blue]{hyperref}

\usepackage{amssymb}
\usepackage[section] {placeins}
\usepackage{epsfig}
\usepackage{graphicx}
\usepackage{color}
\usepackage{dcolumn}
\usepackage{bm}
\usepackage{mdframed}
\usepackage{marginnote}
\usepackage{float}
\usepackage{lineno}
\usepackage{longtable}
\usepackage{rotating}
\usepackage{array}
\usepackage{multicol}
\usepackage{multirow}
\usepackage{xcolor}
\usepackage{tikz}
\usepackage{multirow}
\usepackage{mathtools}

\usepackage[english]{babel}
\usepackage[T1]{fontenc}
\usepackage[utf8]{inputenc}

\usepackage{booktabs} 
\usepackage{lipsum}

\usepackage{mathtools, cuted}

\newcommand{\br}{{\mathbf r}}
\setcitestyle{square}

\DeclarePairedDelimiter{\evdel}{\langle}{\rangle}
\newcommand{\ex}{\evdel}

\journalname{Eur. Phys. J. C}
\begin{document}\sloppy

\title{R--Matrix Calculations for Few--Quark Bound States}

 
 \author{M. A. Shalchi\thanksref{addr1}
        \and
        M. R. Hadizadeh\thanksref{addr2,addr3}
}

\institute{Instituto de F\'{\i}sica Te\'orica, UNESP, 01140-070, S\~ao Paulo, SP, Brazil\label{addr1}
\and
Institute of Nuclear and Particle Physics and Department of Physics and Astronomy, Ohio University, Athens, OH 45701, USA\label{addr2}
 \and 
College of Science and Engineering, Central State University, Wilberforce, OH 45384, USA.\label{addr3}
}

\date{Received: date / Accepted: date}

\maketitle

\begin{abstract}
The R--matrix method is implemented to study the heavy charm and bottom diquark, triquark, tetraquark and pentaquarks in configuration space, as the bound states of quark--antiquark, diquark--quark, diquark--antidiquark and diquark--antitriquark systems, respectively. 
The mass spectrum and the size of these systems are calculated for different partial wave channels. The calculated masses are compared with recent theoretical results obtained by other methods in momentum and configuration spaces and also by available experimental data.
\end{abstract}
\section{Introduction}

The original idea of the R--matrix theory was introduced by Kapur and Peierls \cite{Kapur}, to remove unsatisfactory reliance on perturbation theory in nuclear reactions. It was a few years later that 
Wigner simplified the idea to the formulation of R--matrix theory in which all expressions are energy dependent \cite{Wigner-Physrev-70,Wigner-Physrev-70p,Wigner-phys72}.
Although the R--matrix theory was originally developed for treatment of nuclear resonances \cite{Wigner-phys72,Lane-Mod}, it can also be used to describe all types of reaction phenomena and can be considered as an elegant method to solve the Schr\"{o}dinger equation. These extensions especially became possible after the work of Bloch \cite{Bloch}, by introducing a singular operator between internal and external regions.

The solution of the Schr\"{o}dinger equation for the bound states of few--quarks in momentum space is numerically difficult to handle, because the confining part of the potential leads to singularity at small momenta. To overcome this problem we have successfully used a regularized form of the quark--antiquark \cite{Hadizadeh_AIP1296} and recently diquark--antidiquark ($D\bar D$) \cite{Hadizadeh-PLB753,Hadizadeh-submitted} interactions to solve the Lippmann--Schwinger (LS) equation in momentum space and calculate the mass spectrum of heavy quarkonia and tetraquarks.
To this aim one can keep the divergent part of the potential fixed after exceeding a certain distance, called the regularization cutoff. This procedure creates an artificial barrier and the influence of tunneling barrier is manifested by significant changes in the energy
eigenvalues at small distances.

Recently we have shown that the homogeneous LS equation can be formulated in configuration space to study the heavy tetraquarks as a bound state of $D\bar D$ system \cite{Hadizadeh-EPJC75}.
The variational methods are also used to study the few--quark bound states \cite{Park_NPA925, Kamimura_PRA38, Brink_PRD57, Brac_ZPC, Vijande_PRD76, Vijande_EPJA19}. In order to obtain the variational energy, one must minimize the lowest eigenvalue with respect to the variational parameters after diagonalizing the Hamiltonian matrix. The variational energy can be obtained by differentiating the lowest eigenvalue with respect to the variational parameters.
The successful application of the R--matrix theory to describe the resonance and scattering states resulting from the interaction of particles or systems of particles, which can be nucleons, nuclei,
electrons, atoms and molecules \cite{Burke_2011,Descouvemont_RPP73}, motivated us to implement it to the few--quark bound states.
In this paper we have shown that R--matrix theory is an effective and efficient method to 
study the bound states of few--quarks by solving the Schr\"{o}dinger equation in different partial wave channels.
The successful implementation of the R--matrix method to heavy few--quark systems, paves the path to accurately predict the masses of few--quark bound states composed of light quarks.

\section{R--matrix method for few--quark bound states in configuration space}

In our study for diquark, triquark, tetraquark and pentaquark systems, we have used a two--body picture by considering them as the bound states of quark--antiquark ($q\bar{q}$), diquark--quark ($Dq$), diquark--antidiquark ($D\bar{D}$) and diquark--antitriquark ($D\bar{T}$) systems, respectively. 
The nonrelativistic bound state of any of these two--body systems with the pair relative distance $\br$ in a partial wave representation can be described by the Schr\"{o}dinger equation. 
For simplicity of the notation we use $AB$ for representation of these systems, where $A$ and $B$ stand for any of the subsystems.
Since the $AB$ interaction $V(r)$ is a central force, thus the wave function of few-quark bound states consists of some type of radial function times a spherical harmonic function, i.e. $\Psi_{nlm}(\br) = R_{nl}(r) Y_{lm}(\theta,\phi)$.
The radial function can be obtained by solution of the differential equation
\begin{eqnarray} \label{eq.Schrodinger}
&& \left ( H_{l}-E \right)R_{nl}(r) = \cr
&& \hskip-.5cm \left[-\dfrac{\hbar ^2}{2\mu} \Biggl (\dfrac{d^2}{dr^2}+\dfrac{2}{r}\dfrac{d}{dr}-\dfrac{l(l+1)}{r^2} \Biggr)+V(r)-E\right]R_{nl}(r)=0, \cr
\end{eqnarray}
where $E=m_{AB}-m_A - m_{B}$ is $AB$ binding energy ($m_{AB}$ is the mass of $AB$ system composed of two subsystems $A$ and $B$ with masses $m_A$ and $m_B$). $\mu = \dfrac{m_{A} m_{B}}{m_A +m_{B}}$ is the reduced mass and $l$ is the orbital angular momentum of relative motion of $AB$ system.
In R--matrix calculations, the radial wave function is considered in two internal ($r<r_c$) and external ($r>r_c$) regions as $R_{int}$ and $R_{ext}$, correspondingly. The parameter $a$ is large enough to be sure that the mass spectrum is independent of it.

 The internal wave function $R_{int}$ is defined as combination of the basis functions $u_i(r)$
\begin{equation} \label{Rintdirect}
 R_{int}=\sum_{i=1}^{N} c_i \, u_i(r),
\end{equation}
where
\begin{equation}
 u_i(r)=r^l \, e^{\left(-\dfrac{r^2}{2\rho_i^2} \right)}.
\end{equation}
In our study, the basis states parameters $\rho_i$ are
\begin{equation}
 \rho_i=\rho_0\lambda^{i-1}, \quad ~\rho_0=1.2\dfrac{r_c}{\lambda^{N-1}},\quad \lambda=1.3,\quad N=30.
\end{equation}

Since in the external region the interaction between diquark and antidiquark is fixed, the external wave function can be considered as modified spherical Bessel function of the second kind
\begin{equation}
R_{ext}=C_l \, k_l(\kappa r),
\end{equation}
where $C_l$ is a constant parameter and $\kappa=\sqrt{2 \mu\, (-E+V_0)}$.
Continuity of the internal and external wave functions and derivatives implies that
\begin{eqnarray} \label{Continuity}
 R_{int}(r_c) &=& R_{ext}(r_c), \cr
 R'_{int}(r_c)&=&R'_{ext}(r_c).
\end{eqnarray}

The Hamiltonian $H_{l}$ is not Hermitian in the internal region. To
avoid this the Bloch operator is defined as
\begin{equation}
 L(b)=\dfrac{\hbar^2}{2\mu} \, \delta \left(r-r_c \right) \, \left(\dfrac{d}{dr}-\dfrac{b}{r} \right).
\end{equation}
The dimensionless parameter $b$ is an arbitrary real constant. The delta function indicates that the Bloch operator is a surface operator acting only on $r = r_c$. The operator $H_{l} + L(b)$ is Hermitian, and therefore has a discrete spectrum in the finite region.
Using Eq. (\ref{Continuity}), the Schr\"{o}dinger equation in the internal region can be approximated by
\begin{equation}\label{eq.sch-bloch}
 \biggl ( H_{l}+L(b)-E \biggr ) \, R_{int}(r)=L(b) \, R_{ext}(r).
\end{equation}
It means the logarithmic derivative of the wave function is continuous at $r=r_c$.
By multiplying Eq. (\ref{eq.sch-bloch}) with $u_{i'}$
and integrating in the internal region, 
we obtain the following equation to determine the unknown coefficients $c_i$
\begin{eqnarray}\label{eq2}
 \sum_{i=1}^N \, C_{i'i}(E,b) \, c_i &=& 
 \langle u_{i'}|L(b)|R_{ext}\rangle \cr
 &=&
 \dfrac{\hbar^2 r_c}{2\mu} \, u_{i'}(r_c) \, \biggl ( aR'_{ext}(r_c)-bR_{ext}(r_c) \biggr),
\end{eqnarray}
where
\begin{equation} \label{eq.C}
 C_{i'i}(E,b)=\left \langle u_{i'} \left | H_{l}+L(b)-E \right | u_i \right \rangle.
\end{equation}

Solving Eq. (\ref{eq2}) for $c_i$ and substituting them into Eq. (\ref{Continuity}), i.e $R_{int}(r_c)=\sum_{i=1}^N c_i\, u_i(r_c) = R_{ext}(r_c)$, leads to
\begin{equation}
 R_{int}(r_c)=R(E,b) \, \biggl ( a\,R'_{ext}(r_c)-b\,R_{ext}(r_c) \biggr),
\end{equation}
where $R(E, b)$ is the R--matrix given by
\begin{equation} \label{eq.Reb}
 R(E,b)=\dfrac{\hbar^2 r_c}{2\mu} \sum_{i,i'=1}^N u_{i}(r_c) \, C^{-1}_{ii'}(E,b) \, u_{i'}(r_c).
\end{equation}

The wave function in the internal region is then given by
\begin{equation}\label{eq.Rint}
 R_{int}(r)=\dfrac{\hbar^2 r_c}{2\mu \, R(E,b)} \, R_{ext}(r_c) \sum_{i,i'=1}^N u_{i}(r) \, C^{-1}_{ii'}(E,b) \, u_{i'}(r_c).
\end{equation}

By choosing $b = \dfrac{a \, R'_{ext}(r_c)}{R_{ext}(r_c)} = \dfrac{\kappa r_c \, k'_l(\kappa r_c)}{k_l(k r_c)}$, where $k'_l(\kappa r_c) = \dfrac{dk_l}{d(\kappa r_c)}$, the right hand side of Eq. (\ref{eq2}) will be zero and consequently leads to the following Schr\"{o}dinger-Bloch equation
\begin{equation}
 \sum_{i=1}^{N} \left \langle u_{i'} \left |H_{l}+L \biggl (b(E) \biggr)-E \right | u_i \right \rangle c_i=0.
\label{eq.SB}
\end{equation}
Since $b$ depends on $\kappa$ or the $AB$ binding energy $E$ that we want to calculate, we have written $b \equiv b(E)$.
The equation (\ref{eq.SB}) can be written schematically as eigenvalue equation 
\begin{equation} \label{eq.eigenvalue}
{\cal A}\, c = \lambda \, {\cal B} \, c
\end{equation}
where the matrix elements of ${\cal A}$ and ${\cal B}$ matrices can be obtained as
\begin{eqnarray}
 {\cal A}_{ij} &=& \left \langle u_{i} \left |H_{l}+L \biggl (b(E) \biggr) \right | u_j \right \rangle , \cr
 {\cal B}_{ij} &=& \left \langle u_{i} \left | \right. u_i \right \rangle.
\label{eq.AB-matrix}
\end{eqnarray}
Since the ${\cal A}$ matrix is energy dependent, the solution of the eigenvalue equation (\ref{eq.eigenvalue}) can be started by an initial guess for the energy $E$ and the search in the binding energy can be stopped when $\left |\dfrac{\lambda - E}{E} \right| \le 10^{-10}$.
In order to solve the Eq. (\ref{eq.SB}), we have discretized the continuous variable $r$ with Gauss-Legendre points using a hyperbolic--linear mapping \cite{Hadizadeh-PLB753}. To this aim we have transferred $[0,\infty)$ domain to $[0,1]\cup[1,2]\cup[2,15]\, \text{GeV}^{-1}$ using $75, 75$ and $50$ nodes in each subinterval, respectively.
It indicates that the parameter $r_c$, which divides the configuration space into internal and external regions, is chosen to be $r_c=15\, \text{GeV}^{-1}$ and we have numerically verified that the calculated masses of tetraquarks are independent of the regularization cutoff $a$. 

By having the $AB$ binding energy and eigenvector $c$ from the solution of eigenvalue Eq. (\ref{eq.eigenvalue}), we can calculate the $AB$ internal wave function by equations 
(\ref{eq.C}), (\ref{eq.Reb}) and (\ref{eq.Rint}). Of course, one can calculate the internal wave function directly using Eq. (\ref{Rintdirect}).
Using $AB$ wave function, we can evaluate the expectation value $\ex{r}$ for $AB$ pair distance as
\begin{eqnarray}
\ex{r} &=& \int_0^{\infty} dr\, r^3 R^2(r) \cr 
&=& \int_0^{r_c} dr\, r^3 R_{int}^2(r) +  \int_{r_c}^{\infty} dr\, r^3 R_{ext}^2(r),
\end{eqnarray}
where the $AB$ radial wave function is normalized to 1, i.e. $\int_0^{\infty} dr\, r^2 R^2(r) = 1$.

\section{Results and Discussion} \label{results}

For numerical solution of the integral equation (\ref{eq.Schrodinger}) for $q \bar q$, $Dq$ and $D \bar D$ we have used the spin-independent interaction
\begin{equation}
 \label{eq.V_234}
V (r) = V_{Coul} (r) + V_{conf} (r),
\end{equation}
with the linear confining
\begin{equation}
 \label{eq.V2}
V_{conf} (r) = a \,r+b,
\end{equation}
and the Coulomb-like one-gluon exchange potential 
\begin{equation}
 \label{eq.V2}
V_{Coul} (r) = \gamma \, \dfrac{F_{A} (r) F_{B} (r)}{r}, \quad \gamma = \dfrac{-4}{3} \alpha_s.
\end{equation}
$F_A$ and $F_{B}$ are the form factors of the subsystems $A$ and $B$, correspondingly, and have the following functional form
\begin{equation}
 \label{eq.F}
F (r)= 1-e^{\alpha r -\beta r^2}.
\end{equation}
The parameters of this model are fixed from the analysis of heavy quarkonia masses and radiative decays \cite{Galkin-SJNP44,Galkin-SJNP51,Galkin-SJNP55}.

\subsection{Heavy quarkonia}

For this first test of application of R--matrix method, we have solved the integral equation (\ref{eq.Schrodinger}) to calculate the mass spectra of heavy quarkonia, mesons consisting heavy quark and antiquark. We have used the linear confining plus coulomb potential of Eq. (\ref{eq.V_234}) with form factor $F(r)=1$.
The parameters of potentials are $a=0.18$ GeV$^2$, $b=-0.29$ GeV with $\alpha_s=0.47$ for charmonium ($m_c=1.56$ GeV) and $\alpha_s=0.39$ for bottomonium ($m_b=4.93$ GeV). 
As we have shown in Table \ref{Table_qbarq}, our numerical results for masses of charmonium and bottomonium, obtained by R--matrix method are in excellent agreement with solution of Lippmann--Schwinger integral equation in momentum \cite{Hadizadeh_AIP1296} and configuration \cite{Faustov-IJMPA15} spaces and also with the experimental data \cite{Barnett-PRD54}.

\begin{table}[hbt] 
\caption{The mass spectra of charmonium $\psi(c\bar{c})$ and bottomonium $\Upsilon(b \bar b)$ for the linear confining plus coulomb potential of Eq. (\ref{eq.V_234}) with form factor $F(r)=1$. The masses are given in GeV. The numbers in parentheses are the expectation value of the relative distance between quark--antiquark pair $\ex{r}$ in units of fm.
}
\centering
\begin{tabular}{ccccccccccccccc} 
\toprule
\multicolumn{1}{c}{State} && \multicolumn{4}{c}{$\psi(c\bar{c})$} &&  \multicolumn{4}{c}{$\Upsilon(b \bar b)$} \\
 \cline{3-6}   \cline{8-11} 
 && R--matrix &  LS \cite{Hadizadeh_AIP1296}  & Faustov et al. \cite{Faustov-IJMPA15} & Exp. \cite{Barnett-PRD54}
 && R--matrix &  LS \cite{Hadizadeh_AIP1296}  & Faustov et al. \cite{Faustov-IJMPA15} & Exp. \cite{Barnett-PRD54}  \\ \hline
$1s$ && 3.062 (0.349)  & 3.062 & 3.068  &    3.0675 &&  9.421 (0.184) &    9.425 & 9.447  &   9.4604 \\
$1p$ && 3.529 (0.599) &  3.529 & 3.526  &    3.525  &&   9.910 (0.368) &    9.909 & 9.900   &   9.900 \\
$2s$ && 3.696 (0.734)  &  3.696 & 3.697  &   3.663   && 10.005 (0.453) &    10.006 & 10.012 &   10.023 \\
$1d$ && 3.832 (0.795) &  3.832 & 3.829  &    3.770  &&   10.158 (0.511) &    10.158 & 10.155 &    \\
$2p$ && 3.997 (0.920)  &  3.997 & 3.993  &              &&   10.263 (0.594) &    10.263 & 10.260 &  10.260 \\
$3s$ && 4.144 (1.040) &  4.144 & 4.144  &   4.159   &&  10.349 (0.669) &    10.350 & 10.353 &   10.355 \\
$2d$ && 4.238 (1.081) &  4.237 & 4.234  &               &&  10.451 (0.711) &  10.450 & 10.448 &    \\
$3p$ && 4.387 (1.222) &  4.384 & 4.383  &               &&   10.547 (0.786) &    10.546 & 10.544 &   \\
\bottomrule
\end{tabular}
\label{Table_qbarq}
\end{table}

\subsection{Heavy baryons}

In the next step we have calculated the masses of the ground state heavy baryons consisting of two light $(u; d; s)$ and one heavy $(c; b)$
quarks in the heavy--quark--light--diquark approximation. 
The used diquark mass and form factor parameters are given in Table \ref{Table.pot_parameters}.
As we have shown in Table \ref{Table.baryon}, our numerical results for different heavy baryons, calculated by nonrelativistic R--matrix method, are in good agreement with relativistic and spin-dependent results of EFG~\cite{Ebert_PRD72} and also with MLW results \cite{Mathur_PRD66} of lattice nonrelativistic QCD.
The relative difference between our and Ebert, Faustov, and Galkin (EFG) group results is less than $1.2~ (4.8) \, \%$ for bottom (charm) baryons. Clearly the difference comes from the relativistic effects and also spin terms of the potential that we have ignored in our calculations.

\begin{table}[hbt] 
\caption{The mass $m$ and form factor parameters $\alpha$ and $\beta$ of light--light and heavy--light diquarks. $q$ stands for up and down quarks, and $S$ and $A$ denote the scalar and axial vector diquarks.}
\centering
\begin{tabular}{cccccccccccc}
\toprule
quark  & Diquark  &  $m$ &$\alpha$ &$\beta$       \\ 
content & type & (GeV) & (GeV) & (GeV$^2$)
\\ \hline
\multirow{2}{*}{$qq$} & $S$ & 0.710 & 1.09 & 0.185  \\ \cmidrule{2-5}  
& $A$ & 0.909 & 1.185 & 0.365  \\ 
\hline
\multirow{2}{*}{$qs$} & $S$ & 0.948 & 1.23 & 0.225  \\ \cmidrule{2-5}  
& $A$ & 1.069 & 1.15 & 0.325  \\ 
\hline
$ss$ & $A$ & 1.203 & 1.13 & 0.280  \\ 
\hline
\multirow{2}{*}{$cq$} & $S$ & 1973& 2.55 &0.63 \\ \cmidrule{2-5}  
& $A$ & 2.036& 2.51 &0.45 \\
\hline
\multirow{2}{*}{$cs$}& $S$ & 2091& 2.15 & 1.05 \\  \cmidrule{2-5} 
 & $A$ & 2.158&2.12& 0.99 \\ 
\hline
\multirow{2}{*}{$bq$}& $S$ &  5359 &6.10 & 0.55 \\ \cmidrule{2-5} 
& $A$ & 5.381& 6.05 &0.35 \\
\hline
\multirow{2}{*}{$bs$}& $S$ & 5462 & 5.70 &0.35 \\ \cmidrule{2-5} 
& $A$ & 5.482 & 5.65 &0.27 \\
\bottomrule
\end{tabular}
\label{Table.pot_parameters}
\end{table}

\begin{table}[hbt]
\caption{Masses of the ground states of heavy charm and bottom baryons in units of GeV, calculated by nonrelativistic R--matrix method compared to results of EFG (from solution of relativistic LS equation), to MLW results (obtained from lattice nonrelativistic QCD) and also to the experimental date. The numbers in parentheses are the expectation value of the relative distance between light diquark and heavy quark pair $\ex{r}$ in units of fm. The symbols $[]$ and $\{\}$ denote the scalar and axial vector diquarks.}
\centering
\begin{tabular}{cccccccccccc}   
\toprule
Baryon & content & $I(J^P)$ & R--matrix & EFG~\cite{Ebert_PRD72} & MLW~\cite{Mathur_PRD66} & EXP PDG~\cite{Eidelman_PLB592}  \\ \hline
$\Lambda_c$ & $\{ss\}c$  & $0(\frac12^+)$ & 2.396 (0.49) &  2.297 & 2.290  & 2.2849(0.006)   \\
$\Sigma_c$ & $\{uu\}c$ & $1(\frac12^+)$ & 2.554 (0.45) & 2.439 & 2.452  & 2.4513(0.007)  \\
$\Xi_c$ & $[us]c$ & $\frac12(\frac12^+)$& 2.594 (0.45) &  2.481 & 2.473  & 2.4663(0.0014)  \\
$\Xi'_c$ & $\{us\}c$ & $\frac12(\frac12^+)$& 2.702 (0.44) & 2.578 &  2.599 & 2.5741(0.0033)\\ 
$\Omega_c$ &  $\{ss\}c$  & $0(\frac12^+)$& 2.829 (0.43) &  2.698 & 2.678 & 2.6975(0.0026)\\
$\Lambda_b$ & $[ud]b$ & $0(\frac12^+)$& 5.687 (0.45) &  5.622 & 5.672  & 5.6240(0.009)   \\
$\Sigma_b$ & $\{uu\}b$ & $1(\frac12^+)$ & 5.843 (0.42) & 5.805 & 5.847 &  \\
$\Xi_b$ & $[us]b$ & $\frac12(\frac12^+)$& 5.882 (0.41) &  5.812 &  5.788 &\\
$\Xi'_b$ & $\{us\}b$ & $\frac12(\frac12^+)$& 5.989 (0.40) & 5.937 &  5.936 &\\
$\Omega_b$ & $\{ss\}b$ & $0(\frac12^+)$& 6.115 (0.39) & 6.065 &  6.040 &\\
\bottomrule
\end{tabular}
\label{Table.baryon}
\end{table}

\subsection{Heavy tetraquarks}

In our calculations for heavy tetraquarks we have used the masses of diquark (antidiquark) and form factor parameters of Ref. \cite{ebert2006masses} which are given in Table \ref{Table.pot_parameters}.

Our numerical results for the masses of charm ($cq\bar{c}\bar{q}$ and $cs\bar{c}\bar{s}$) and bottom ($bq\bar b\bar{q}$ and $bs\bar b\bar{s}$) tetraquarks 
for $s-$, $p-$ and $d-$wave channels with total spin ${\cal S}=0$ are listed in Tables \ref{Table.cqcq_cscs} and \ref{Table.bqbq_bsbs}. The tetraquark masses are calculated for scalar $S\bar{S}$ and axial-vector $A\bar{A}$ diquark--antidiquark contents.
We have also calculated the expectation value of the relative distance between $D\bar D$ pair which can provide an estimate of the size of the tetraquarks.

\begin{table}[hbt]
\caption{Masses of charm diquark--antidiquark states in units of GeV, calculated by nonrelativistic R--matrix method compared to nonrelativistic Lippmann--Schwinger (LS) calculations and relativistic EFG results. The numbers in parentheses are the expectation value of the relative distance between $D\bar D$ pair $\ex{r}$ in units of fm.}
\centering
\begin{tabular}{ccccccccccccccc}   
\toprule
state & \multicolumn{4}{c}{$cq\bar{c}\bar{q}\,(S\bar{S})$} & & \multicolumn{4}{c}{$cq\bar{c}\bar{q}\,(A\bar{A})$} \\ \cline{2-5} \cline{7-10} 
 & R--matrix & LS \cite{Hadizadeh-PLB753} & EFG~\cite{ebert2006masses,Ebert_EPJC58} & LS \cite{Hadizadeh-EPJC75} && R--matrix & LS \cite{Hadizadeh-PLB753} & EFG~\cite{ebert2006masses,Ebert_EPJC58}  & LS \cite{Hadizadeh-EPJC75} \\ \hline
$1s$ & 3.885 (0.35) & 3.792 & 3.812 & 3.885 && 4.013 (0.35) & 3.919 & 3.852 & 4.013 \\
$1p$  & 4.268 (0.55) & 4.262 & 4.244 & && 4.388 (0.54) & 4.374 &4.350 \\
$2s$  & 4.461 (0.70) & 4.419 & 4.375 & && 4.580 (0.69) & 4.535 & 4.434 \\
$1d$  & 4.553 (0.73) & 4.556 & 4.506 & && 4.669 (0.72) & 4.668 & 4.617 \\
$2p$  & 4.708 (0.85) & 4.697 & 4.666 & && 4.823 (0.84) & 4.816 & 4.765 \\
$3s$  & 4.873 (0.99) & 4.843  & & && 4.988 (0.97) & 4.944  \\
$2d$  & 4.932 (1.00) & 4.933 & & && 5.044 (0.99) & 5.037  \\
$3p$  & 5.073 (1.10) & 5.062 & & && 5.184 (1.09) & 5.184  \\
\hline
state & \multicolumn{4}{c}{$cs\bar{c}\bar{s}\,(S\bar{S})$} & & \multicolumn{4}{c}{$cs\bar{c}\bar{s}\,(A\bar{A})$} \\ \cline{2-5} \cline{7-10}
 & R--matrix & LS \cite{Hadizadeh-PLB753} & EFG~\cite{ebert2006masses,Ebert_EPJC58} & LS \cite{Hadizadeh-EPJC75}  && R--matrix & LS \cite{Hadizadeh-PLB753} & EFG \cite{ebert2006masses,Ebert_EPJC58} & LS \cite{Hadizadeh-EPJC75}  \\ \hline
 $1s$  & 4.117 (0.35) & 4.011  & 4.051 & 4.117 && 4.249 (0.34) & 4.139  & 4.110 & 4.250 \\
$1p$  & 4.490 (0.54) & 4.490  & 4.466 & && 4.617 (0.53) & 4.616 & 4.582  \\
$2s$  & 4.681 (0.68) & 4.620  & 4.604 & && 4.808 (0.68)  & 4.744 & 4.680 \\
$1d$  & 4.770 (0.71) & 4.770 & 4.728 & && 4.893 (0.71) & 4.894 & 4.847 \\
$2p$  & 4.922 (0.83) & 4.920  & 4.884 & && 5.045 (0.82) & 5.041  & 4.991 \\
$3s$  & 5.086 (0.95) & 5.039 & &&& 5.208 (0.95) & 5.160 \\
$2d$  & 5.142 (0.98) & 5.143 & &&& 5.262 (0.97) & 5.263  \\
$3p$  & 5.281 (1.08) & 5.276 & &&& 5.399 (1.07) & 5.394  \\
\bottomrule
\end{tabular}
\label{Table.cqcq_cscs}
\end{table}

\begin{table}[hbt]
\caption{The same as Table \ref{Table.cqcq_cscs}, but for bottom tetraquarks.}
\centering
\begin{tabular}{cccccccccccccccc}   
\toprule
state & \multicolumn{3}{c}{$bq\bar b \bar q\,(S\bar{S})$} & & \multicolumn{3}{c}{$bq\bar b \bar q\,(A\bar{A})$} \\ \cline{2-4} \cline{6-8}
 & R--matrix & LS \cite{Hadizadeh-PLB753} & EFG~\cite{ebert2006masses,Ebert_MPLA24} & & R--matrix & LS \cite{Hadizadeh-PLB753} & EFG~\cite{ebert2006masses,Ebert_MPLA24} \\ \hline
$1s$  & 10.482 (0.23) & 10.426 & 10.471 & & 10.527 (0.23) & 10.469  & 10.473 \\
$1p$  & 10.814 (0.38) & 10.813  & 10.807  && 10.858 (0.38) & 10.856  & 10.850 \\
$2s$  & 10.942 (0.47) & 10.914  & 10.917  && 10.986 (0.48) & 10.958  & 10.942 \\
$1d$  & 11.034 (0.51) & 11.034 &  11.021 && 11.077 (0.51) & 11.077 &  11.064  \\
$2p$  & 11.142 (0.59) & 11.140  & 11.122 && 11.185 (0.59) & 11.183 & 11.163 \\
$3s$  & 11.252 (0.68) & 11.230 & && 11.295 (0.68) & 11.273  \\
$2d$  & 11.311 (0.70) & 11.310 & && 11.354 (0.70) & 11.354  \\
$3p$  & 11.409 (0.78) & 11.406 & && 11.452 (0.78) & 11.450   \\
\hline
state & \multicolumn{3}{c}{$bs\bar b \bar s\,(S\bar{S})$} & & \multicolumn{3}{c}{$bs\bar b \bar s\,(A\bar{A})$} \\ \cline{2-4} \cline{6-8}
 & R--matrix & LS \cite{Hadizadeh-PLB753} & EFG~\cite{ebert2006masses,Ebert_MPLA24} & & R--matrix & LS \cite{Hadizadeh-PLB753} & EFG~\cite{ebert2006masses,Ebert_MPLA24} \\ \hline
 $1s$ & 10.691 (0.23) & 10.629 & 10.662  && 10.732 (0.23) & 10.668  & 10.671 \\
$1p$ & 11.017 (0.38) & 11.015 & 11.002  && 11.056 (0.38) & 11.054  & 11.039  \\
$2s$ & 11.146 (0.48) & 11.116 & 11.111 &&  11.186 (0.48) & 11.155  & 11.133  \\
$1d$ & 11.235 (0.51) & 11.235 &  11.216  && 11.275 (0.51) & 11.274 & 11.255  \\
$2p$ & 11.343 (0.59) & 11.340  & 11.316  && 11.382 (0.59) & 11.379 & 11.353  \\
$3s$ & 11.454 (0.68) & 11.430 & && 11.493 (0.68) & 11.469  \\
$2d$ & 11.511 (0.70) & 11.511 & && 11.550 (0.70) & 11.549  \\
$3p$  & 11.608 (0.78) & 11.606 & && 11.647 (0.77) & 11.645 \\
\bottomrule
\end{tabular}
\label{Table.bqbq_bsbs}
\end{table}

We have compared our results for tetraquark masses with recent results obtained in momentum and configuration spaces by solution of the nonrelativistic Lippmann--Schwinger integral equation \cite{Hadizadeh-PLB753,Hadizadeh-EPJC75} and also with those of previous relativistic studies by EFG reported in Refs. \cite{ebert2006masses,Ebert_EPJC58,Ebert_MPLA24}. 

It indicates that the calculated masses for the ground state of charm tetraquarks (i.e. $cq\bar{c}\bar{q}$ and $cs\bar{c}\bar{s}$) with corresponding results from LS (in configuration space) \cite{Hadizadeh-EPJC75} shows that they are in excellent agreement.
Our results are also in good agreement with those of LS (in momentum space) and EFG with a relative percentage difference estimated to be at most $2.5 \,\%$ and $4\,\%$, respectively.
While our R--matrix calculations and LS results are both done in a nonrelativistic spin--independent scheme, there is some difference between the results obtained in momentum and configuration spaces. 
Clearly, the difference between our R--matrix results in configuration space and EFG results in momentum space is larger, and comes from the relativistic effects and also spin contribution in the $D\bar{D}$ interaction which appears in spin--orbit, spin--spin and tensor spin--space terms \cite{Ebert_EPJC58}.
As we have shown in Ref. \cite{Hadizadeh-PLB753} the relativistic effect leads to a small reduction in the mass of heavy tetraquarks and decreases the masses of charm and bottom tetraquarks by less than $2\,\%$ and $0.2\,\%$, respectively, whereas the spin contribution may lead to small decrese or increase in the masses of tetraquarks.

\begin{figure}[hbt]
 \caption{The $x-z$ cross section of the $s-$, $p-$ and $d-$wave probability densities $|\Psi_{nlm}(\br)|^2$ for $bq\bar b\bar{q}$ tetraquark in $S\bar{S}$ state for $m=0$.} 
 \centering
 \includegraphics[width=.8\textwidth]{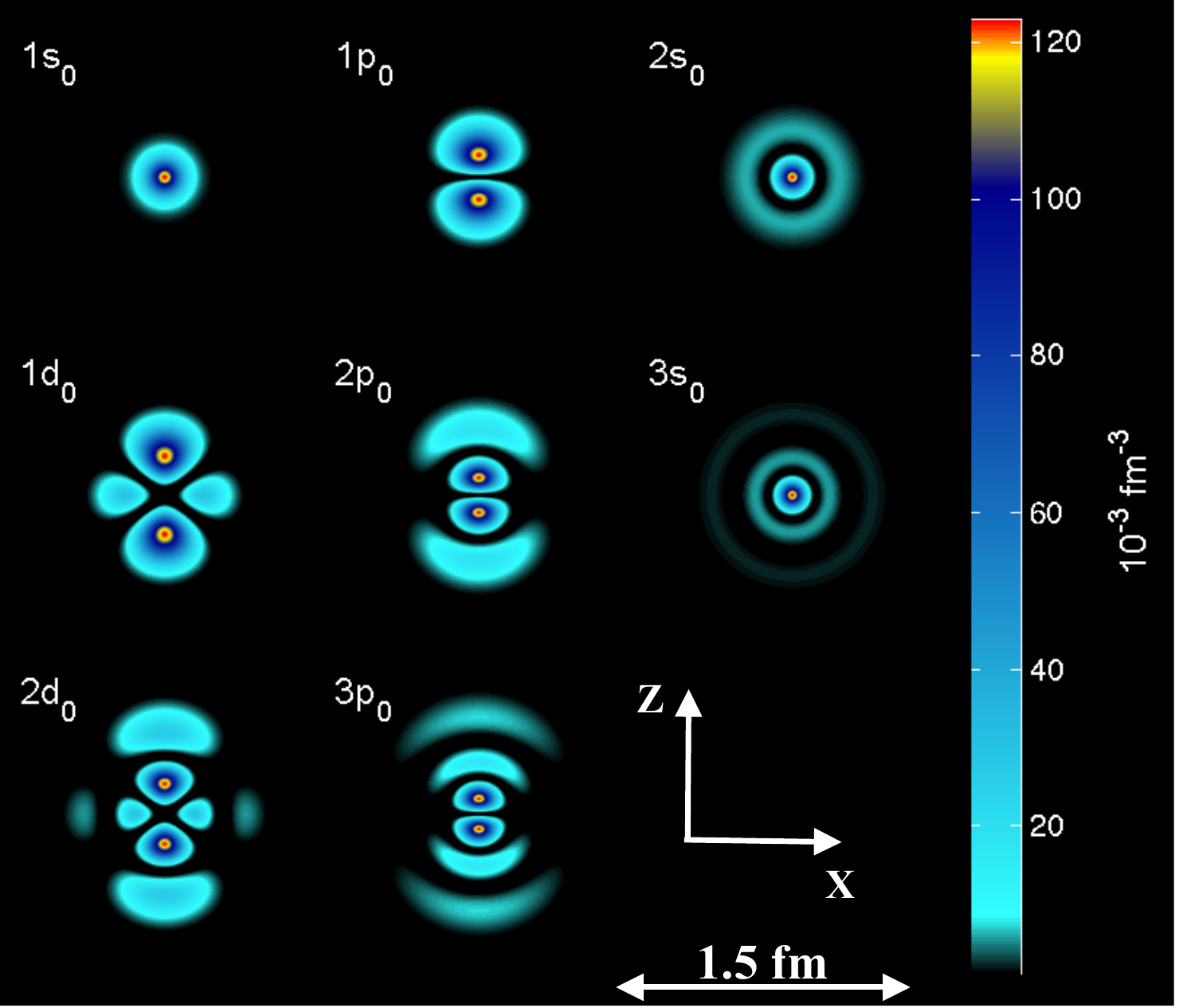}
 \label{Fig.densities}
\end{figure}

In Fig. \ref{Fig.densities} we have shown few examples of the probability density $|\Psi_{nlm}(\br)|^2$ of $bq\bar b\bar{q}$ tetraquark in $S\bar{S}$ state for $s$, $p$ and $d$ channels.
As we can see the tetraquark probability functions for higher states have been expanded to larger distances which leads to larger expectation value for the relative distance between $D \bar D$ pair. 
As we can see in Tables \ref{Table.cqcq_cscs} and \ref{Table.bqbq_bsbs}, the expectation value of the relative distance between $D \bar D$ pair is almost the same for scalar $S\bar{S}$ and axial-vector $A\bar{A}$ diquark--antidiquark contents.
It is larger for higher states and its size changes roughly with a factor of 3 from $1s$ to $3p$ state.
In Table \ref{Table.cqcq_experiment}, we have compared our results for the masses of charm and bottom tetraquarks with the possible experimental candidates. They are in good agreement with a relative difference below $3.4\,\%$.

\begin{table}[hbt] 
\caption{Comparison of our numerical results for the masses of charm and bottom diquark--antidiquark states, calculated by R--matrix method and possible experimental candidates. The expectation value $\ex{r}$ for diquark--antidiquark pair distance is also calculated. }
\centering
\begin{tabular}{cccccccccccccc}  
\toprule
& \multirow{2}{*}{state} & \multicolumn{2}{c}{R--matrix Theory} & & \multicolumn{2}{c}{Experiment} \\ \cline{3-4} \cline{6-7}
&  & Mass (MeV) & $\ex{r}\, (\text{fm})$ & & Exp. candidate & Mass (MeV)     \\ \hline 
 \multirow{9}{*}{\rotatebox[origin=c]{90}{$cq\bar{c}\bar{q}\,(S\bar{S})$}} \\
& & & & & & 
 \multirow{3}{*}{
$
\left\{
\begin{array}{l l}
 4259 \pm 8^{+2}_{-6} & \text{\cite{PhysRevLett.95.142001}} \\
 4247 \pm 12^{+17}_{-32} & \text{\cite{PhysRevLett.99.182004}} \\
 4284 ^{+17}_{-16} \pm 4 & \text{\cite{He-PRD74}} \\
\end{array}
\right.
$ 
} \\ 
& $1p$ & 4268 & 0.5500 & & $Y(4260)$ &\\
& & & & & \\
\\
& & & & & & 
 \multirow{3}{*}{
 $
\left\{
\begin{array}{l l}
 4664\pm11\pm5 & \text{\cite{PhysRevLett.99.142002}} \\\\
 4634^{+8+5}_{-7-8} & \text{\cite{PhysRevLett.101.172001}}
\end{array}
\right.
$
} \\ 
& $2p$ & 4708 &0.8491 & & $Y(4660)$ &\\
& & & & & \\ \\ 
\hline
 \multirow{9}{*}{\rotatebox[origin=c]{90}{$cq\bar{c}\bar{q}\,(A\bar{A})$}} \\
& & & & & & 
 \multirow{3}{*}{$
\left\{
\begin{array}{l l}
 4361\pm9\pm9 & \text{\cite{PhysRevLett.99.142002}} \\
 4324\pm24 & \text{\cite{PhysRevLett.98.212001}} \\
 4355^{+9}_{-10} \pm 9 & \text{\cite{{Liu_PRD78}}}   
\end{array}
\right.
$} \\ 
& $1p$ & 4388 & 0.5449 && $Y(4360)$ &\\
& & & & & \\
\\
& & & & & & 
 \multirow{3}{*}{$
\begin{array}{l l}
 4433\pm4\pm2 & \text{\cite{PhysRevLett.100.142001} }
\end{array}
$} \\ 
& $2s$ & 4580 &0.6924 && $Z(4430)$ &\\
& & & & & \\ 
\hline
 \multirow{9}{*}{\rotatebox[origin=c]{90}{$bq\bar b\bar{q}\,(S\bar{S})$}} \\
 & & & & & &
 \multirow{3}{*}{$
\left\{
\begin{array}{l l}
 10876 \pm 2  & \text{\cite{Aubert_PRL102}} \\ \\
 10865 \pm 8 & \text{\cite{Yao_JPG33}}   
\end{array}
\right.
$} 
\\
& $1p$ & 10814 & 0.3780 && $Y(10860)$   \\  \\ \\
 & & & & & &
 \multirow{3}{*}{$
\left\{
\begin{array}{l l}
 10996 \pm 2  & \text{\cite{Aubert_PRL102}} \\ \\
 11019 \pm 8 & \text{\cite{Yao_JPG33}}   
\end{array}
\right.
$} 
\\
& $2p$ & 11142 & 0.5948 && $Y(11020)$   \\ \\
\bottomrule
\end{tabular}
\label{Table.cqcq_experiment}
\end{table}

\subsection{Pentaquarks}

\begin{figure}[hbt] 
 \caption{Pentaquark system as a diquark--antitriquark bound state.} 
\centering
 \includegraphics[width=.65\textwidth]{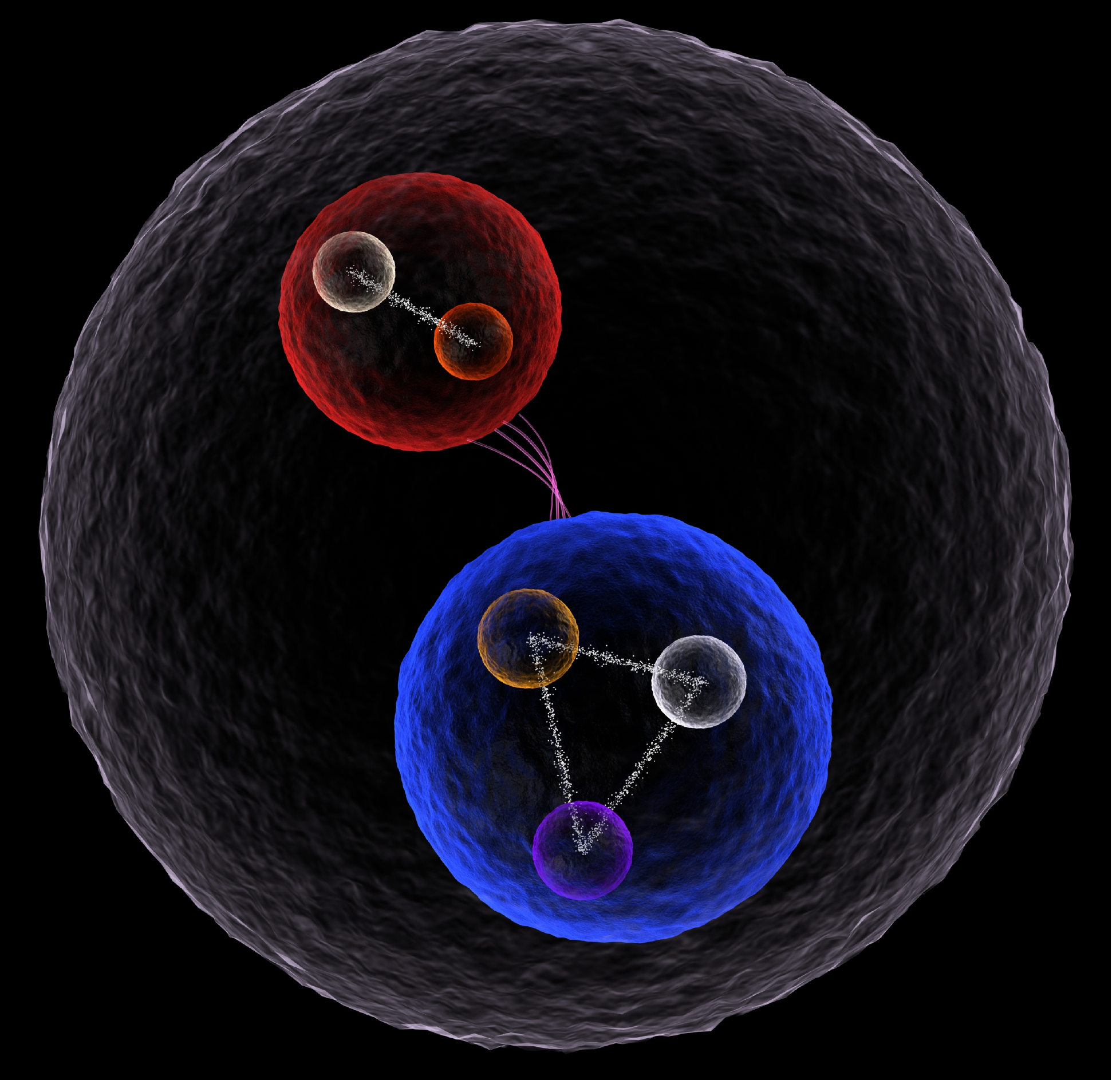}    
 \label{Fig.pentaquark}
\end{figure}

By successful application of the R--matrix method for diquark, triquarks and tetraquarks we have also implemented it to study the pentaquarks as bound states of diquark--antitriquark systems (see Fig. \ref{Fig.pentaquark}).

In this study we have used two models of $D\bar T$ interaction. In the following sections we present the models and our numerical results.

\subsubsection{One--pion exchange potential}

One--pion exchange potential (OPEP) acting between a nucleon and a heavy meson ($D$ or $B$) given as \cite{Cohen_PRD72}
\begin{eqnarray}
V_{\pi} (r) =\left\{
                \begin{array}{ll}
                 V_0 ,  & r< r_0
                  \\ \\
                 2 {\bf I}_N \cdot {\bf I}_H  \biggl ( S_{12} V_T(r) + 2 {\bf S}_N \cdot {\bf S}_l V_c(r)  \biggr ) , & r > r_0 
                \end{array}
              \right.
\end{eqnarray}
where
\begin{eqnarray}
V_0 &=&  -62.79 \quad \text{or} \quad  -276 \, \text{MeV} , \cr
V_T(r) &=& \dfrac{g_A g_H m_{\pi}^2 }{2 \pi f_{\pi}^2}  \dfrac{e^{-m_{\pi}r}}{6r}  \left ( \dfrac{3}{m_{\pi}^2 r^2} + \dfrac{3}{m_{\pi}r} +1 \right ), \cr
V_c(r) &=& \dfrac{g_A g_H m_{\pi}^2 }{2 \pi f_{\pi}^2}  \dfrac{e^{-m_{\pi}r}}{3r}  , 
\end{eqnarray}
and the labels are:
nucleon isospin $I_N$, heavy-meson isospin $I_H$ (total isospin ${\bf I} = {\bf I}_N + {\bf I}_H$), tensor force $S_{12}$, tensor potential $V_T (r)$, nucleon spin $S_N$, light quark in heavy meson spin $S_l$ (sum of nucleon spin and light quark spin ${\bf K} = {\bf S}_N + {\bf S}_l$) and central potential $V_c(r)$.
The parameters of OPEP potential are given in Table \ref{Table.OPEP}.

\begin{table}[hbt] 
\caption{The parameters of OPEP potential used in the heavy pentaquark calculations.}
\centering
\begin{tabular}{lllllllll}
\toprule
parameter &&& value \cr
\hline
$g_A$ &&& $1.27$ \cr
$f_{\pi}$ &&& $131$ MeV \cr
$g_H$ &&& $+0.59 $ \cr
$m_{\pi}$ &&& $138$ MeV \cr
$m_N$ &&& $938.92$ MeV \cr
$m_B$ &&& $5279$ MeV \cr
$m_D$ &&& $1867$ MeV \cr
\bottomrule
\end{tabular}
\label{Table.OPEP}
\end{table}

Our results for the masses of the pentaquarks composed of a nucleon and a meson ($B$ and $D$ mesons) for $l=0$, $S=\frac{3}{2}$ and $J^P=\frac{3}{2}^+$ channel are given in Table \ref{tab:B-D_meson}. As we can see our results with R--matrix method are in excellent agreement with Cohen et al. results \cite{Cohen_PRD72}.

\begin{table}[hbt]
\caption{The masses of pentaquarks composed of a nucleon and a meson for $l=0$, $S=\frac{3}{2}$ and $J^P=\frac{3}{2}^+$ channel. Upper panel: nucleon + $B$ meson, lower panel: nucleon plus $D$ meson. All masses are in GeV\@.  Column 
A: constant potential,
$V_0=-276$~MeV and $r_0 = 1$~fm; 
B: constant potential,
$V_0=-62.79$~MeV and $r_0 = 1.5$~fm.}
\centering
\begin{tabular}{lllllllll} 
\toprule
& \multicolumn{2}{c}{A} && \multicolumn{2}{c}{B} \\
\cmidrule{2-3} \cmidrule{5-6} 
 & \cite{Cohen_PRD72} & R--matrix && \cite{Cohen_PRD72} & R--matrix  \\ 
 \hline
 $I=0$ &  6.077  & 6.076  && 6.202 & 6.201  \\
 $I=1$ &  6.077 & 6.076 && 6.203 & 6.202  \\  
 \hline
 $I=0$ & 2.689 & 2.688 && 2.797 & 2.795   \\ 
$I=1$ & 2.691 & 2.689  && 2.797 & 2.796  \\ 
\bottomrule
\end{tabular}
\label{tab:B-D_meson}
\end{table}

\subsubsection{Cornell Potential}
For the second test of Pentaquark calculations, the nonrelativistic linear--plus--Coulomb Cornell potential is used. It has the following form \cite{Barnes_PRD72}

\begin{equation} \label{Cornell_penta}
V(r) = V_c(r) + V_{spin-spin}+ V_{spin-orbit} + V_{tensor}
\end{equation}
where
\begin{eqnarray}
V_c(r) &=& -\dfrac{4}{3} \dfrac{\alpha_s}{r} + br \cr
V_{spin-spin} &=& \dfrac{32 \pi \alpha_s}{9 m_D m_T} \left ( \dfrac{\sigma}{\sqrt{\pi}}  \right )^3  e^{-\sigma^2 r^2} {\bf S}_D \cdot  {\bf S}_{\bar T} \cr 
V_{spin-orbit} &=& \dfrac{1}{m_D m_T} \left (  \dfrac{2\alpha_s}{r^3} - \dfrac{b}{2r} \right ) {\bf L} \cdot  {\bf S} \cr 
V_{tensor} &=& \dfrac{1}{m_D m_T} \dfrac{4\alpha_s}{r^3} T,
\end{eqnarray}
and spin--spin, spin--orbit and tensor operators can be calculated as
\begin{eqnarray}
{\bf S}_D \cdot  {\bf S}_{\bar T} &=& \dfrac{1}{2} \left ( {\bf S}^2 - {\bf S}_D^2 - {\bf S}_{\bar T}^2  \right ), \cr
{\bf L} \cdot  {\bf S} &=& \dfrac{1}{2} \left ( {\bf J}^2 - {\bf L}^2 - {\bf S}^2  \right ), \cr
\langle ^3L_J | T | ^3L_J \rangle &=& 
 \left\{
                \begin{array}{ll}
                  \dfrac{-L}{6(2L+3)},  & J=L+1
                  \\  
                 \dfrac{1}{6}, & J=L
                  \\
                  -\dfrac{L+1}{6(2L-1)},  & J=L-1.
                \end{array}
              \right.
\end{eqnarray}
In Table \ref{Table.cornel-pentaquark}, the parameter of Cornell potential used in pentaquark calculations are given.

\begin{table}[hbt] 
\caption{The parameters of Cornell potential used in heavy pentaquark calculations.}
\centering
\begin{tabular}{lllllllll}
\toprule
parameter &&& value \cr
\hline
$\alpha_s$ &&& $0.5461$ \cr
$b$ &&& $0.1425$ GeV$^2$ \cr
$\sigma$ &&& $1.0946$ GeV \cr
$m_D$ &&& $1.860$ GeV \cr
$m_{\bar T}$ &&& $2.286$ GeV \cr
\bottomrule
\end{tabular}
\label{Table.cornel-pentaquark}
\end{table}

Our results for the masses of charmoniumlike pentaquark $P_c^+$ in $s-$ and $p-$wave channels are given in Table \ref{Table.Pc+}.
In our calculations we have ignored tensor force and we have considered central, spin--spin and spin--orbit terms of diquark--antitriquark interaction of Eq. (\ref{Cornell_penta}).
Beside the small difference between our results for the masses of $s-$ and $p-$wave charmoniumlike pentaquarks and Lebed's results \cite{Lebed_PLB749}, which is about $5 \%$ and comes from neglected tensor force, the diquark--antitriquark separation $\ex{r}$ for $s-$wave channel with value of $0.306$ fm is almost half of the obtained separation by Lebed with value of $0.64$ fm. For the $p-$wave channel, neglecting the tensor force leads to the relative difference of $14\%$ in calculated separations.

\begin{table*}[hbt] 
\caption{Masses of charmoniumlike pentaquark $P_c^+$ in a diquark--antitriquark picture in units of MeV, calculated by nonrelativistic R--matrix method compared to Lebed's results. The diquark--antitriquark separation $\ex{r}$ is also calculated in units of fm.}
\centering
\begin{tabular*}{\textwidth}{@{\extracolsep{\fill}}lllllllllllllllll@{}}
\toprule
state & $J^P$ & potential & Mass (MeV) & $\ex{r}$ (fm)   \\ \hline
$s-$wave   & $\frac{3}{2}^-$  \\
& & & \multicolumn{2}{c}{R--matrix} \\  \cmidrule{4-5}  
                  &  & $V_c$  & 4112 &  0.292 \\
                  &  & $V_c + V_{spin-spin} (+ V_{spin-orbit} ) $  & 4151 &  0.306  \\
 &  &  &  \multicolumn{2}{c}{Theory \cite{Lebed_PLB749}} \\ \cmidrule{4-5} 
  &  & $V(r)$ & $4380$ & $0.64$  \\ 
   &  &  &  \multicolumn{2}{c}{\cite{Aaij_PRL115}} \\ \cmidrule{4-5} 
  &  & Experimental candidate $P_c^+(4380)$ & $4380\pm8\pm29$  \\ 
  \hline
$p-$wave   & $\frac{5}{2}^-$   \\
& & & \multicolumn{2}{c}{R--matrix} \\  \cmidrule{4-5}  
                  &  & $V_c$  & 4593 & 0.554  \\
                  &  & $V_c + V_{spin-spin}$  & 4597 & 0.559   \\
                  &  & $V_c + V_{spin-spin} + V_{spin-orbit} $ & 4633 &  0.600   \\
                   &  &  &  \multicolumn{2}{c}{Theory \cite{Lebed_PLB749}} \\ \cmidrule{4-5} 
  &  & $V(r)$ & $4450$ & $0.70$  \\
     &  &  &  \multicolumn{2}{c}{\cite{Aaij_PRL115}} \\ \cmidrule{4-5} 
  &  & Experimental candidate $P_c^+(4450)$ & $4449.8\pm1.7\pm2.5$  \\ 
\bottomrule
\end{tabular*}
\label{Table.Pc+}
\end{table*}

In conclusion, we have implemented the R--matrix method to calculate the mass spectra of heavy quarkonia, baryons, tetraquarks and pentaquarks in the two-body picture. 
Our numerical results for the masses of heavy charm and bottom few--quarks even by neglecting the relativistic effects, are in good agreement with other theoretical predictions and also with available experimental data.

\begin{acknowledgements}
We would like to thank Y. Suzuki for helpful discussions about the application of R--matrix method to two--body problems, to R. F. Lebed for helpful comments on Cornel potential used in Pentaquark calculations and to Reza Abedi, 3D modeler and character artist, for creating the pentaquark picture.
M.A.S. acknowledges the financial support from the Brazilian agency CAPES and M.R.H. acknowledges the partial support by National Science Foundation under Contract No. NSF-HRD-1436702 with Central State University and by the Institute of Nuclear and Particle Physics at Ohio University.

\end{acknowledgements}

\end{document}